\documentclass[11pt,a4wide]{article}
\usepackage{newlfont}
\usepackage{amsmath}
\usepackage{graphics}
\usepackage{float}
\usepackage{graphicx}
\usepackage{epsfig}
\usepackage{multirow}
\begin{document}

\title{Masses, Radii and Regge Trajectories of $\Sigma_{u}^{-}$ State Hybrid Charmonium}
\author{Nosheen Akbar\thanks{e mail: nosheenakbar@cuilahore.edu.pk,noshinakbar@yahoo.com} \\
\textit{$\ast$Department of Physics, COMSATS University Islamabad, Lahore Campus,}\\
{Lahore(54000), Pakistan.}}
\date{}
\maketitle

\begin{abstract}
In this paper, masses and radii of $\Sigma^-_u$ states hybrid charmonium mesons are calculated by numerically solving the Schr$\textrm{\"{o}}$dinger equation with non-relativistic potential model. Results for calculated masses of $\Sigma^-_{u}$ states charmonium hybrid mesons are found to be close to the results obtained through lattice simulations. Calculated masses are used to construct Regge trajectories. It is found that the trajectories are almost linear and parallel.
\end{abstract}



\section*{I. INTRODUCTION}

Study of charmonium mesons is very important in particle physics. Conventional charmonium mesons, that can be described by the quark model, are the charm quark--antiquark pairs bounded with ground state gluonic field; while hybrid charmonium mesons are the quark--antiquark pairs with excited state gluonic field. In \cite{morning, morningstar03}, quarkonium hybrid mesons are discussed through lattice simulations and potentials for different states of quarkonium hybrids are plotted in fig.3 of \cite{morningstar03}. These states are characterized by quantum numbers, $J, \, L, \, S$, $\Lambda$, $\eta$, and $\epsilon$, where $J = L\oplus S$ with $L$ as the orbital angular momentum quantum number and $S$ as the spin angular momentum quantum number. $\Lambda$ is the projection of the total angular momentum of gluons and for $\Lambda = 0, \pm1, \pm2, \pm3, .... $, meson states are represented as $\Sigma,\Pi,\Delta$ and so on \cite{morning}. $\eta$ is the combination of parity and charge and for $\eta = P \circ C = +,-$, states are labelled by sub-script $g,u$ \cite{morning}. $\epsilon$ is the eigen value corresponding to the operator $P$ and is equal to $+,-$. Parity and charge for hybrid static potentials are defined as \cite{morning}
\begin{equation}
P=\epsilon (-1)^{L+\Lambda +1},C= \epsilon \eta (-1)^{L+\Lambda +S},
\label{cp}
\end{equation}

The low-lying static potential states are labelled as $\Sigma_g^{+}, \Sigma_g^{-},\Sigma_u^{+},\Sigma_u^{-}, \Pi_{g}, \Pi_{u}, \Delta_{g}, \Delta_{u}$ and so on\cite{morning}. $\Sigma_g^{+}$ is the low-lying potential state with ground state gluonic field and is approximated by a coulomb plus linear potential. The $\Pi_u$ and $\Sigma^-_u$ are the $Q\overline{Q}$ potential states with low lying gluonic excitations. Linear plus coulombic potential model is extended in \cite{Nosheen11} for $\Pi_u$ states by fitting the suggested ansatz with lattice data\cite{morningstar03} and the extended model is tested by finding properties of mesons for a variety of $J^{PC}$ states in refs.\cite{Nosheen11,Nosheen14, Nosheen17,Nosheen19}. In ref.\cite{Nosheen20}, linear plus coulombic potential model is extended for lowest excited hybrid state, $\Sigma^{-}_{u}$ by fitting the lattice data \cite{morning} with the suggested analytical expression (ansatz). The validity of suggested ansatz is tested by calculating the spectrum of $\Sigma^{-}_{u}$ state bottomonium mesons. In this paper, this extended potential model\cite{Nosheen20} is used to calculate the spectrum and radii of the of $\Sigma^{-}_{u}$ charmonium hybrid mesons. For this purpose, Born-Oppenheimer formalism and adiabatic approximation is used. Using these calculated masses of $\Sigma^-_u$ states hybrid charmonium mesons, their Regge trajectories are constructed.

A Regge trajectory is a graph of mass square versus the angular momentum quantum number (or the principle quantum number) of a particular family of particles. Different potential models give different Regge trajectories as in \cite{9906293},\cite{1603.04725},\cite{0910.5612},\cite{godfrey}.
In 1985, Godfrey and Isgur \cite{godfrey} solved the Schrodinger equation with the incorporation of spin-spin and spin-orbit interactions and obtained linear Regge trajectories. Linear Regge trajectories are also observed in \cite{0910.5612}, \cite{godfrey}, \cite{0903.5183}, \cite{0408124} while non-linear Regge trajectories are observed in ref.\cite{9906293,Ghosh}. Regge trajectories for charmonium and bottomonium mesons with spinless Salpeter-type equation by employing the Bohr-Sommerfeld quantization approach are presented by Chen in ref.\cite{jiao2018}.

This paper is organised as: In the section 2 of this paper, Potential model for $\Sigma^-_u$ is defined while methodology of calculation of mass and radii is explained in section 3 and 4. Regge trajectories are constructed  in section 5 whose properties like linearity and parallelism are observed, while the discussion on results and concluding remarks are written in section 6.

\subsection*{2. POTENTIAL MODEL FOR $\Sigma^-_u$ STATES}
The potential model \cite{barnes05} for conventional mesons is extended in ref.\cite{Nosheen20} for the study of $\Sigma^-_u$ bottomonium mesons. This potential model for $\Sigma^-_u$ charmonium states can be written as:
\begin{equation}
V(r)= \frac{-4\alpha _{s}}{3r}+ br + \frac{32\pi \alpha_s}{9 m_c m_{\overline{c}}} (\frac{\sigma}{\sqrt{\pi}})^3 e^{-\sigma ^{2}r^{2}} \textbf{S}_{c}. \textbf{S}_{\overline{c}} + \frac{4 \alpha _{s}}{m^2_c  r^3} S_T + \frac{1}{m_c^2} \big(\frac{2\alpha _{s}}{r^3}- \frac{c}{2r}\big)\textbf{L}.\textbf{S} + A^{\prime} exp (- B^{\prime} r^{P^{\prime}}) + C^{\prime}.
\end{equation}

The terms $\frac{-4\alpha _{s}}{3r}$, $b r$, $\textbf{S}_{c}. \textbf{S}_{\overline{c}}$, $\textbf{L}.\textbf{S}$, and $S_T$ are defined in \cite{Nosheen14},\cite{Nosheen20},\cite{barnes05}.
Spin-orbit and colour tensor terms are equal to zero~\cite{barnes05} for $L=0$. Constituent mass of charm quark, $m_{c}$, is taken to be 1.454 GeV \cite{Nkorean}. The parameters ($\alpha_s, \, b$, $\sigma$) used in this potential model are taken from \cite{Nkorean} and equal to
  0.5315, $ 0.1583 \, \text{ GeV}^{2}$, and 1.105 GeV respectively. The parameters ($A^{\prime}, \, B^{\prime}, P^{\prime}, \, C^{\prime}$) are found in ref.\cite{Nosheen20} by fitting $A^{\prime} exp (- B^{\prime} r^{P^{\prime}}) + C^{\prime}$ with the lattice data \cite{morningstar03} and fitted parameters are:
  
   $A^{\prime}=11.5917 \textrm{GeV}, \quad B^{\prime}=4.6119, \quad {P^{\prime}}= 0.2810, \quad C^{\prime}=0.9589$.

\subsection*{3. MASS OF $\Sigma^-_u$ STATES}

For $\Sigma_u^-$ states, radial Schr$\textrm{\"{o}}$dinger equation can be written as \cite{Nosheen20}:
\begin{equation}
U^{\prime \prime }(r) + 2\mu \left( E - V(r) - \frac{L (L + 1) - 2 \Lambda
^{2}+\left\langle J_{g}^{2}\right\rangle}{2 \mu r^{2}}\right) U(r) = 0,
\end{equation}
Here, V(r) is the potential defined above in eq. (2), $U(r)=r R(r)$, where $R(r)$ is the radial wave function. $\left\langle J_{g}^{2}\right\rangle $ is the square of gluon angular momentum and $\left\langle J_{g}^{2}\right\rangle =2$ \cite{morning} for $\Sigma^-_u$ state. $\Lambda$ is the projection of gluon angular momentum and $\Lambda =0$ \cite{morning} for $\Sigma^-_u$ state.
Numerical solutions of the Schr$\ddot{\textrm{o}}$dinger equation for $\Sigma^-_u$ states are found by the shooting method. At small distance (r $\rightarrow$ 0), wave function becomes unstable due to very strong attractive potential. This problem is solved by applying smearing of position co-ordinates by using the method discussed in ref. \cite{godfrey}.

Masses of $\Sigma_u^{-}$ state charmonium mesons are calculated by adding the constituent quark masses  to the energy $E$ , i.e;
\begin{equation}
M_{c\overline{c}} = m_{c} + m_{\overline{c}}+ E.
\end{equation}

\subsection*{4. RADII}
 To find the root mean square radii of the gluonic excited $\Sigma_u^-$ charmonium states, following relation is used:
\begin{equation}
\sqrt{\langle r^{2}\rangle} = \sqrt{\int U^{\star} r^{2} U dr}.\label{P25}
\end{equation}
Calculated masses and radii for $\Sigma_u^-$ states are reported in Table 1.

\begin{table}[tbp]
\caption{Calculated masses and radii of $c\overline{c}$ hybrid $\Sigma^-_u$ charmonium mesons.}
\begin{center}
\tabcolsep=4pt \fontsize{9}{11}\selectfont
\begin{tabular}{|c|c|c|c|c|}
\hline
Meson &$J^{PC}$ & mass & mass, \cite{2019} & radii \\ \hline
& & \textrm{GeV} & \textrm{GeV} & fm \\ \hline
$\eta^{h}_{c} (1 ^1S_0)$ &$0^{++}$ & 4.5151 & \multirow{2}{1.4cm}{4.487(5)} & 0.6774 \\
$J/\psi^{h} (1 ^3S_1)$ & $1^{+-}$ & 4.5234 & &0.6878 \\ \hline
$\eta^{h}_{c} (2 ^1S_0)$ & $0^{++}$ & 4.9526 &\multirow{2}{1.4cm}{4.933(9)} &1.0574\\
$J/\psi^{h} (2 ^3S_1)$ & $1^{+-}$ & 4.962 & &1.0684\\ \hline
$\eta^{h}_{c} (3 ^1S_0)$ & $0^{++}$ & 5.3191 & & 1.3805 \\
$J/\psi^{h} (3 ^3S_1)$ & $1^{+-}$ & 5.3287 & &1.391 \\ \hline
$\eta^{h}_{c} (4 ^1S_0)$ & $0^{++}$ & 5.6451 & &1.6702 \\
$J/\psi^{h} (4 ^3S_1)$ &$1^{+-}$ & 5.6544 & &1.6801 \\ \hline
$\eta^{h}_{c} (5 ^1S_0)$ & $0^{++}$ & 5.9437 & &1.9369 \\
$J/\psi^{h} (5 ^3S_1)$ & $1^{+-}$ & 5.9528 & &1.9463 \\ \hline
$\eta^{h}_{c} (6 ^1S_0)$ & $0^{++}$ & 6.2223 & &2.1866 \\
$J/\psi^{h} (6 ^3S_1)$ & $1^{+-}$ & 6.2311 & & 2.1954 \\ \hline
$h^{h}_{c} (1 ^1P_1) $ & $1^{--}$ & 4.6856 &\multirow{4}{1.4cm}{4.623(6)} & 0.8008 \\
$\chi^{c}_{0} (1 ^3P_0)$ &$0^{-+}$ & 4.6263 & &0.7607 \\
$\chi^{h}_{1} (1 ^3P_1)$ & $1^{-+}$ & 4.6785 & & 0.7951 \\
$\chi^{h}_{2} (1 ^3P_2)$ & $2^{-+}$ & 4.7044 & &0.8192 \\ \hline
$h^{h}_{c} (2 ^1P_1) $ &$1^{--}$ & 5.0871 & \multirow{4}{1.4cm}{5.058(10)} &1.1612 \\
$\chi^{h}_{0} (2 ^3P_0)$ & $0^{-+}$ & 5.0578 & & 1.1384 \\
$\chi^{h}_{1} (2 ^3P_1)$ & $1^{-+}$ & 5.0866 & &1.1609 \\
$\chi^{h}_{2} (2 ^3P_2)$ & $2^{-+}$ & 5.0989 & &1.1743 \\ \hline
$h^{h}_{c} (3 ^1P_1) $ & $1^{--}$ & 5.4347 & &1.4725 \\
$\chi^{h}_{0} (3 ^3P_0)$ &$0^{-+}$ & 5.4156 & &1.4583 \\
$\chi^{h}_{1} (3 ^3P_1)$ &$1^{-+}$ &5.4365 & &1.4745 \\
$\chi^{h}_{2} (3 ^3P_2)$ & $2^{-+}$ & 5.4442 & & 1.4835 \\ \hline
$h^{h}_{c} (4 ^1P_1) $ & $1^{--}$ & 5.7485 & &1.764 \\
$\chi^{h}_{0} (4 ^3P_0)$ & $0^{-+}$ & 5.7347 & & 1.7441 \\
$\chi^{h}_{1} (4 ^3P_1)$ &$1^{-+}$ & 5.7515& &1.7572 \\
$\chi^{h}_{2} (4 ^3P_2)$ & $2^{-+}$ & 5.7569 & & 1.764 \\ \hline
$h^{h}_{c} (5 ^1P_1) $ & $1^{1--}$ & 6.0384& &2.0147 \\
$\chi^{h}_{0} (5 ^3P_0)$ & $0^{-+}$ & 6.028 & &2.0074 \\
$\chi^{h}_{1} (5 ^3P_1)$ & $1^{-+}$ & 6.0422 & &2.0185 \\
$\chi^{h}_{2} (5 ^3P_2)$ & $2^{-+}$ & 6.0462 & &2.024 \\ \hline
$h^{h}_{c} (6 ^1P_1) $ & $1^{--}$ & 6.3104 & &2.2596 \\
$\chi^{h}_{0} (6 ^3P_0)$ & $0^{-+}$ & 6.3023 & & 2.2541 \\
$\chi^{h}_{1} (6 ^3P_1)$ & $1^{-+}$ & 6.3146 & &2.2638 \\
$\chi^{h}_{2} (6 ^3P_2)$ & $2^{-+}$ & 6.3176 & &2.2684 \\ \hline

$\eta^{h}_{c2} (1 ^1D_2)$ & $2^{++}$ & 4.8941 & \multirow{4}{1.4cm}{4.814(7)}&0.9572 \\
$\psi_1^{h} (1 ^3D_1)$ & $1^{+-}$ & 4.8517 & &0.9088 \\
$\psi_2^{h} (1 ^3D_2)$ &$2^{+-}$ & 4.89 & &0.9494 \\
$\psi_3^{h} (1 ^3D_3)$ & $3^{+-}$ & 4.913 & & 0.9846 \\ \hline

$\eta^{h}_{c2} (2 ^1D_2)$ & $2^{++}$ & 5.2608& & 1.2966 \\
$\psi_1^{h} (2 ^3D_1)$ & $1^{+-}$ & 5.2354 & &1.2695 \\
$\psi_2^{h} (2 ^3D_2)$ &$2^{+-}$ & 5.2593 & &1.2934 \\
$\psi_3^{h} (2 ^3D_3)$ & $3^{+-}$ & 5.2736 & &1.3131 \\ \hline

$\eta^{h}_{c2} (3 ^1D_2)$ & $2^{++}$ & 5.5878 & & 1.5951 \\
$\psi_1^{h} (3 ^3D_1)$ & $1^{+-}$ &5.5692 & &1.5759 \\
$\psi_2^{h} (3 ^3D_2)$ &$2^{+-}$ & 5.5874 & &1.5936 \\
$\psi_3^{h} (3 ^3D_3)$ & $3^{+-}$ & 5.5978 & &1.6078 \\ \hline

$\eta^{h}_{c2} (4 ^1D_2)$ & $2^{++}$ & 5.5878 & & 1.8674 \\
$\psi_1^{h} (4 ^3D_1)$ & $1^{+-}$ &5.5692 & &1.8526 \\
$\psi_2^{h} (4 ^3D_2)$ &$2^{+-}$ & 5.5874 & &1.867 \\
$\psi_3^{h} (4 ^3D_3)$ & $3^{+-}$ & 5.5978 & &1.8782 \\ \hline

\end{tabular}
\end{center}
\end{table}

\subsection*{5. REGGE TRAJECTORIES}

Using the calculated masses of charmonium meson (given in Table 1), Regge trajectories are constructed in (J, $M_{c\overline{c}}^2$) and
($n$,$M^2$) planes. To construct the Regge trajectories, particle's states are selected with consecutive values of J. In figs. 1, 2, 3 and 4, Regge trajectories are constructed in (J,$M^2$) planes for $\Sigma^-_u$ state hybrid charmonium mesons by using the following relation \cite{Ghosh}
\begin{equation}
M_{c\overline{c}}^2 = \alpha J + \alpha_0,
\end{equation}
Slope ($\alpha$) and intercept ($\alpha_0$) are found by fitting the above equation with equation of straight line for particular set of particles that lie on a single trajectory. The fitted slopes and intercepts are given in Table 2.
\begin{figure}
\begin{center}
\epsfig{file=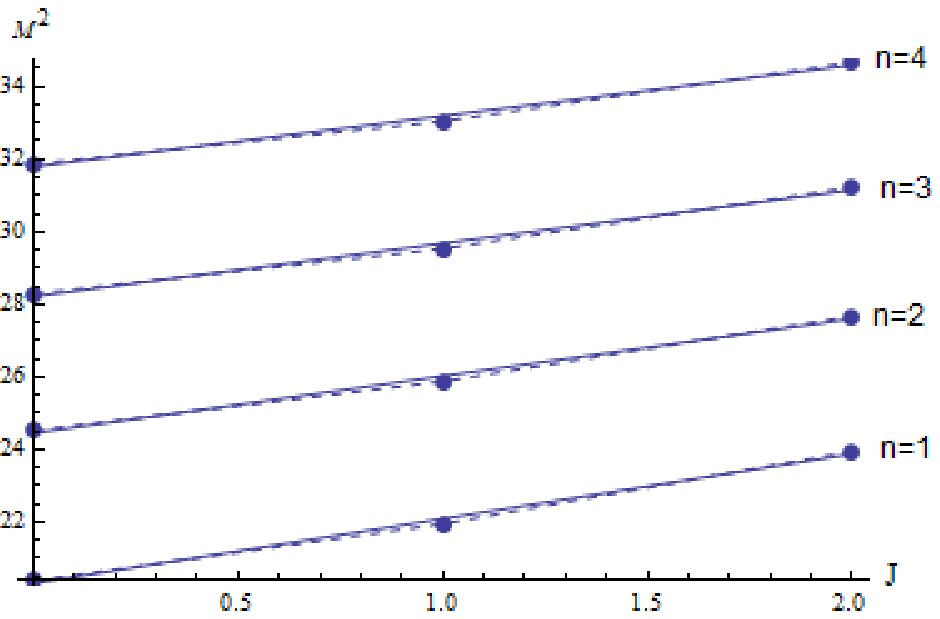,width=0.5\linewidth,clip=}\caption{Regge trajectories for $^1 S_0, \, ^1 P_1, \,^1 D_2$ hybrid charmonium $\Sigma^-_u$ states. Dashed lines are for Regge trajectories constructed with calculated masses (shown by dots). Solid lines are for the Regge trajectories obtained with $M_{c\overline{c}}^2 = \alpha J + \alpha_0$.}
\end{center}
\end{figure}

\begin{figure}
\begin{center}
\epsfig{file=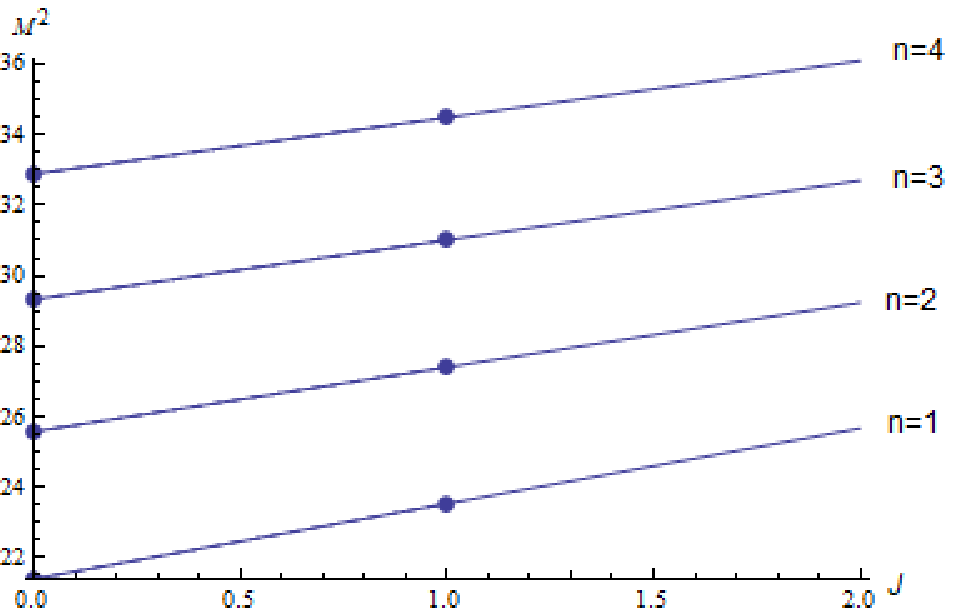,width=0.5\linewidth,clip=}\caption{Regge trajectories for $^3 P_0, \, ^3 D_1$ hybrid charmonium $\Sigma^-_u$ states. Dashed lines are for Regge trajectories constructed with calculated masses (shown by dots). Solid lines are for the Regge trajectories obtained with $M_{c\overline{c}}^2 = \alpha J + \alpha_0$.}
\end{center}
\end{figure}

\begin{figure}
\begin{center}
\epsfig{file=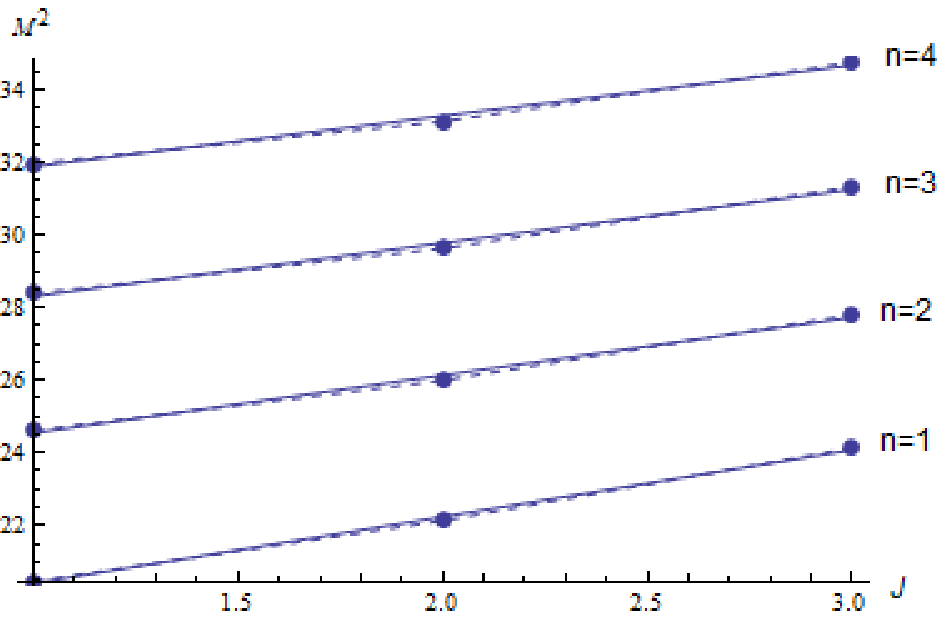,width=0.5\linewidth,clip=}\caption{Regge trajectories for $^3 S_1, \, ^3 P_2, \, ^3 D_3$ hybrid charmonium $\Sigma^-_u$ states. Dashed lines are for Regge trajectories constructed with calculated masses (shown by dots). Solid lines are for the Regge trajectories obtained with $M_{c\overline{c}}^2 = \alpha J + \alpha_0$.}
\end{center}
\end{figure}

\begin{figure}
\begin{center}
\epsfig{file=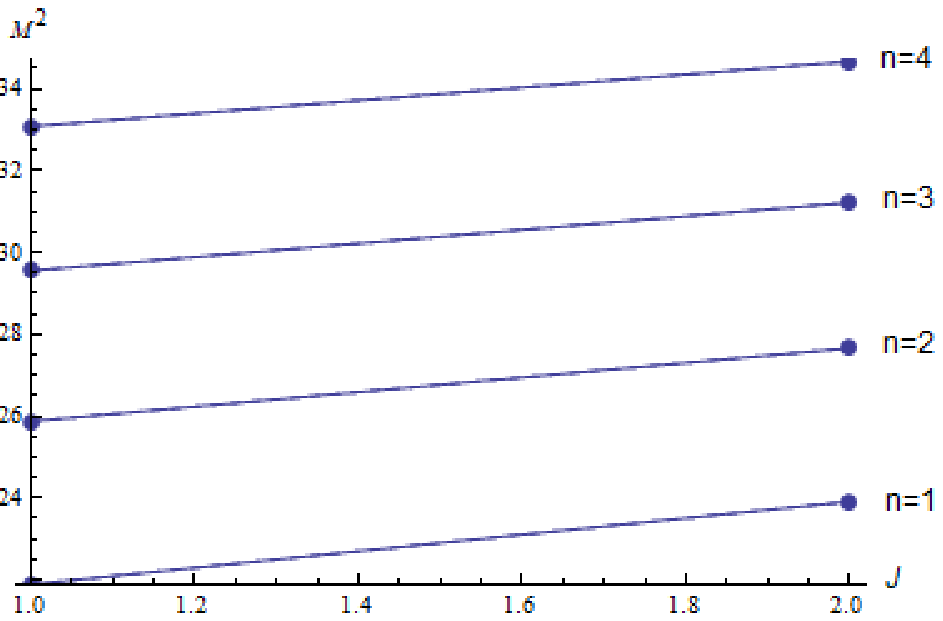,width=0.5\linewidth,clip=}\caption{Regge trajectories for $^3 P_1, \, ^3 D_2$ hybrid charmonium $\Sigma^-_u$ states. Dashed lines are for Regge trajectories constructed with calculated masses (shown by dots). Solid lines are for the Regge trajectories obtained with $M_{c\overline{c}}^2 = \alpha J + \alpha_0$.}
\end{center}
\end{figure}

In figs. 5 Regge trajectories are constructed in (n,$M^2$) plane for $\Sigma^-_u$ state hybrid charmonium by using the following relation:
\begin{equation}
M_{c\overline{c}}^2 = \beta (n - 1) + \beta_0,
\end{equation}
Here, $n$ is the principle quantum number. Slope ($\beta$) and intercept ($\beta_0$) of the above equation are found by fitting it with equation of straight line for particular set of particles that lie on a single trajectory. The fitted slopes and intercepts are given in Table 3.

\begin{figure}
\begin{center}
\epsfig{file=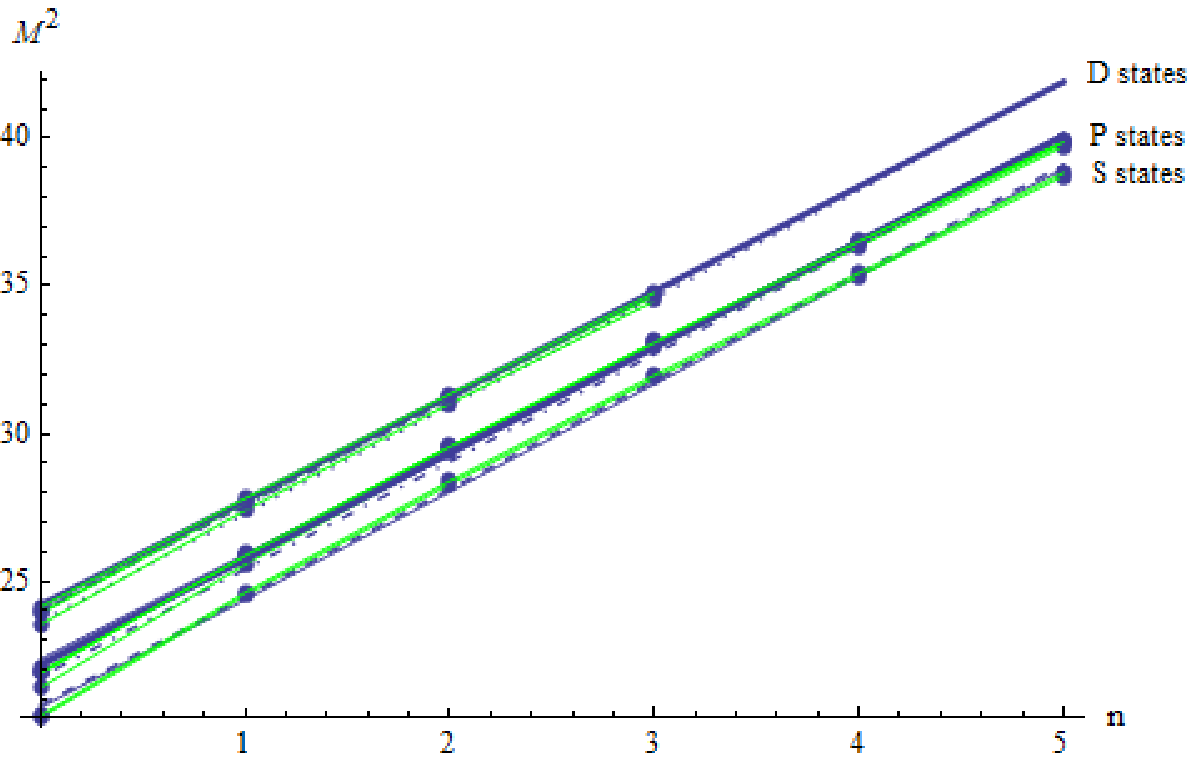,width=0.8\linewidth,clip=}\caption{Regge trajectories for $S, \, P, \, D$ hybrid charmonium $\Sigma^-_u$ states. Solid green color lines are for Regge trajectories constructed with $M_{c\overline{c}}^2 = \beta (n - 1) + \beta_0$. Blue curves are for  Regge trajectories constructed with calculated masses shown by dots. $^1 S_0$ state trajectory is shown by blue solid lines while $^3 S_1$ state rajectory is shown by dashed blue line. $^1 P_1, \, ^3 P_0, \, 3 P_1, \, ^3 P_2$ are shown by dashed, dotted, dot-dashed and thick lines respectively. $^1 D_2, \, ^3 D_1, \, 3 D_2, \, ^3 D_3$ are shown by dashed, dotted, dot-dashed and thick lines respectively.}
\end{center}
\end{figure}
\begin{table}[tbp]
\caption{Parameters $\Sigma_u^-$ states of charmonium meson.}
\begin{center}
\tabcolsep=10pt \fontsize{10}{12}\selectfont
\begin{tabular}{|c|c|c|c|} \hline
& n & $\alpha$ & $\alpha_0$ \\ \hline
\multirow{4}{*}{Fig.1}& 1 & 1.783 & 20.3147\\
 & 2 & 1.5739 & 24.4537 \\
 & 3 & 1.4653 & 28.2188 \\
 & 4 & 1.3983 & 31.7937 \\ \hline

\multirow{4}{*}{Fig.2}& 1 & 2.1363 & 21.4027 \\
 & 2 & 1.8281 & 25.5813 \\
 & 3 & 1.6873 & 29.3287 \\
 & 4 & 1.6053 & 32.8868 \\ \hline

 \multirow{4}{*}{Fig.3}& 1 & 2.0237 & 19.8646 \\
 & 2 & 1.7867 & 24.0868 \\
 & 3 & 1.6635 & 27.892 \\
 & 4 & 1.5888 & 31.491 \\ \hline

\multirow{4}{*}{Fig.4}& 1 & 1.8382 & 18.5669 \\
 & 2 & 1.5947 & 22.9543 \\
 & 3 & 1.4702 & 26.8496 \\
 & 4 & 1.3953 & 30.5017 \\ \hline

\end{tabular}
\end{center}
\end{table}

\begin{table}[tbp]
\caption{Parameters $\Sigma_u^-$ states of charmonium meson.}
\begin{center}
\tabcolsep=10pt \fontsize{10}{12}\selectfont
\begin{tabular}{|c|c|c|c|} \hline
& n & $\beta$ & $\beta_0$ \\ \hline
\multirow{2}{*}{S states}& $^1 S_0$ & 3.6465 & 20.737 \\
 & $^3 S_1$ & 3.6528 & 20.82 \\ \hline

\multirow{4}{*}{P states}& $^1P_1$ & 3.5598 & 22.2169 \\
 & $^3 P_0$ & 3.6402 & 21.7754 \\
 & $^3 P_1$ & 3.5816 & 22.1758 \\
 & $^3 P_2$ & 3.5455 & 22.3667 \\ \hline

\multirow{4}{*}{D states}& $^1 D_2$ & 3.5682 & 24.0265 \\
 & $^3 D_1$ & 3.6466 & 23.6442 \\
 & $^3 D_2$ & 3.5828 & 23.9908 \\
 & $^3 D_3$ & 3.54 & 24.2016 \\ \hline

\end{tabular}
\end{center}
\end{table}

\subsection*{6. DISCUSSION AND CONCLUSIONS}

In this paper, potential model for lowest lying $\Sigma_u^-$ state hybrids is used to calculate the spectrum and radii of $1S-6S$, $1P-6P$, $1D-4D$ hybrid charmonium states and results are written in Table 1. Calculated masses are close to the the results given in ref.\cite{2019} as shown in Table 1. In ref. \cite{2019}, spectrum is calculated without including the spin, so the same mass is given for $\eta^h_c$ and $J/ \psi^h$. However, our proposed potential model gives distinguished results for $S=0$ and $S=1$. As observed from Table 1, the lowest calculated  mass of the $\Sigma_u^-$ state hybrid charmonium is 4.5151 GeV. Difference between the mass obtained in \cite{2019} and the calculated mass is 0.0281 GeV for the lowest state. In ref. \cite{morningstar03}, the lowest mass of $\Sigma_u^-$ charmonium state is $~4.6$ GeV.

From Table 1, it is observed that masses and radii are increased with increase of the orbital quantum number (L) or principle quantum number (n). It shows that mass and radii increase with radial and orbital excitations. The similar behaviour is observed in ref.\cite{Nosheen14} while working on $\Sigma^+_g$ and $\Pi_u$ states of charmonium meson.

It is observed from Table 2 that slope ($\alpha$) decreases and intercept ($\alpha_0$) increases toward higher states. As observed from Table 1 that mass is increased toward higher states, so it can be concluded that slopes decrease (or intercepts increase) with increase of mass. Similar characteristic of the Regge trajectories is observed in \cite{0408124} for charmed strange mesons. The slopes ($\alpha$) of trajectories presented with respect to angular momentum are found to be greater as compared to the slopes ($\beta$) of trajectories presented with respect to principal quantum number. The similar behaviour is observed in ref.\cite{0903.5183} in which slopes of trajectories with the orbital excitations is found to be in average 1.3 times greater than the slope of the trajectories with radial excitation for the light mesons.

In fig.5, $^1 S_0$ and $^3 S_1$ states trajectories are almost same. Similarly for $P$ states with different spin quantum number and $D$ states with different spin quantum numbers, trajectories are almost same. From figs. (1-5), it is observed that the calculated masses of $\Sigma^-_u$ state hybrid charmonium fit to the linear trajectories  presented in ($M^2$, J) and ($M^2$, n) planes. It is also observed that the trajectories are almost parallel. Same characteristics for conventional mesons trajectories are observed in \cite{0903.5183,regge,godfrey}. These Regge trajectories can be helpful for the identification of higher excited states.


\begin{thebibliography}{99}

\bibitem{morning}
K. J. Juge, J. Kuti, and C. Morningstar, J Nucl. Phys. Proc. Suppl. \textbf{63}, 326  (1998).

\bibitem{morningstar03}
K. J. Juge, J. Kuti, and C. Morningstar, AIP Conf. Proc. \textbf{688}, 193 (2003).

\bibitem{Nosheen11}
N. Akbar, B. Masud, and S. Noor, Eur. Phys. J. A $\textbf{47}$, 124 (2011); erratum: Eur. Phys. J. A \textbf{50}, 121 (2014).

\bibitem{Nosheen14}
A. Sultan, N. Akbar, B. Masud, and F. Akram, Phys. Rev. D \textbf{90}, 054001 (2014).

\bibitem{Nosheen17}
N. Akbar, M. A. Sultan, B. Masud, and F. Akram, Phys. Rev. D \textbf{95}, 074018 (2017).

\bibitem{Nosheen19}
N. Akbar, F. Akram, B. Masud, and M. A. Sultan, Eur. Phys. J. A \textbf{55}, 82 (2019).

\bibitem{Nosheen20}
N. Akbar, arxive:2005.02626 (2020).

\bibitem{9906293}
M. M. Brisudova, L. Burakovsky, and T. Goldman, Phys. Rev. D \textbf{61}, 054013 (2000).

\bibitem{1603.04725}
A. M. Badalian, and B. L. G. Bakker, Phys. Rev. D \textbf{93}, 074034 (2016).

\bibitem{0910.5612}
D. Ebert, R. N. Faustov, and V. O. Galkin, Eur. Phys. J. C \textbf{66}, 197 (2010).

\bibitem{godfrey}
S. Godfrey, Phys. Rev. D \textbf{31}, 2375 (1985).

\bibitem{0903.5183}
D. Ebert, R. N. Faustov, and V. O. Galkin, Phys. Rev. D \textbf{79}, 114029 (2009).

\bibitem{0408124}
A.L. Zhang, Phys. Rev. D \textbf{72}, 017902 (2005).

\bibitem{Ghosh}
R. Ghosh, and A. Bhattacharya, Int. J. Theor. Phys. \textbf{56}, 2335 (2017).

\bibitem{jiao2018}
Jiao-Kai Chena, Eur. Phys. J. C  \textbf{78}, 235 (2018).

\bibitem{barnes05}
T. Barnes, S. Godfrey, and E. S. Swanson, Phys. Rev. D \textbf{72}, 054026 (2005).

\bibitem{Nkorean}
N. Akbar, J. Korean Phys. Soc. (2020); arxive:2002.09566 

\bibitem{2019}
S. Capitani, O. Philipsen, C. Reisinger, C. Riehl, and M. Wagner, Phys. Rev. D \textbf{99}, 034502 (2019).

\end{thebibliography}
\end{document}